\documentclass[article, singlecolumn,superscriptaddress,preprintnumbers,nopacs,floatfix,amsmath,amssymb]{revtex4}

\usepackage[english]{babel}

\usepackage{graphicx}
\usepackage{dcolumn}
\usepackage{bm}
\usepackage{amsmath}
\usepackage{amssymb}
\usepackage{appendix}
\usepackage{mathrsfs}
\usepackage{bbm}
\usepackage{epsfig}
\usepackage[mathscr]{euscript}
\usepackage{calligra}
\usepackage[T1]{fontenc}
\usepackage{aurical}
\usepackage[T1]{fontenc}

\usepackage{wrapfig}
\usepackage{eurosym}
\usepackage[usenames,dvipsnames]{color}
\usepackage[yyyymmdd,hhmmss]{datetime}

\def\nn{\nonumber}

\newcommand{\be}{\begin{eqnarray}}
\newcommand{\ee}{\end{eqnarray}}
\newcommand{\br}{\begin{matrix}}
\newcommand{\fr}{\frac}
\newcommand{\pr}{\partial}
\newcommand{\er}{\end{matrix}}

\newcommand{\bt}{\Theta}
\newcommand{\bp}{\Phi}

\newcommand{\ct}{{\vartheta}}
\newcommand{\cp}{{\varphi}}

\newcommand{\vn}{{\vec{n}}}

\newcommand{\acc}{\\[3mm]}

\newcommand{\vct}[1]{\vec{#1}}

\newcommand{\vecn}{\vct{n}}

\newcommand{\vecb}{\vct{b}}
\newcommand{\Vr}{\vct{r}}
\newcommand{\vt}{\vct{t}}

\newcommand{\eastar}{\end{eqnarray*}}

\begin{document}

\title{On relation between discrete Frenet frames and the bi-Hamiltonian \\ structure of the discrete nonlinear Schr\"odinger equation }

\author{Theodora Ioannidou}
\email{
ti3@auth.gr
}
\affiliation{
Faculty of Civil Engineering,  School of Engineering, 
Aristotle University of Thessaloniki, 54249, Thessaloniki, Greece
}
\author{ Antti J. Niemi}
\email{Antti.Niemi@physics.uu.se}
\affiliation{Department of Physics and Astronomy, Uppsala University,
P.O. Box 803, S-75108, Uppsala, Sweden}
\affiliation{Nordita, Stockholm University, Roslagstullsbacken 23, SE-106 91 Stockholm, Sweden}
\affiliation{
Laboratoire de Mathematiques et Physique Theorique
CNRS UMR 6083, F\'ed\'eration Denis Poisson, Universit\'e de Tours,
Parc de Grandmont, F37200, Tours, France}
\affiliation{Department of Physics, Beijing Institute of Technology, Haidian District, Beijing 100081, P. R. China}

\begin{abstract}

The discrete Frenet equation entails  a local framing of a discrete, 
piecewise linear polygonal chain in terms of its bond and torsion angles. In particular, 
the tangent vector of a segment is akin the classical O(3) spin variable. Thus there is a relation to
the lattice Heisenberg model, that can be used to model physical properties of the chain.   On the other hand, the Heisenberg model 
is closely related to the discrete nonlinear Schr\"odinger (DNLS) equation.   Here we apply these
interrelations to develop a perspective on discrete chains dynamics: We employ the properties of a discrete chain in terms of
a  spinorial representation of the discrete Frenet equation, to introduce a bi-hamiltonian structure for 
the discrete nonlinear Schr\"odinger equation (DNLSE),  which we then use to produce integrable chain dynamics.

\end{abstract}


\maketitle

\section{Introduction}

String-like objects have important ramifications in many areas of physics, from understanding turbulent flows in fluid 
dynamics \cite{vortex} to new forms of matter in high energy physics \cite{pauls}.  Similarly, piecewise linear discrete chains have
many applications  from the modelling of proteins in terms of its C$\alpha$ backbone \cite{danielsson,nora1,nora2} to robotics and 3D virtual reality \cite{hansonbook}.

Here, the starting point is the observation by Hasimoto \cite{Has} that the one dimensional nonlinear Schr\"odinger equation (NLSE) describes string-like objects such as vortex filaments. 
For this, a change of variables is introduced that relates the NLSE  to the Frenet frame representation of a space curve. 
The NLSE is a widely studied integrable model with various theoretical \cite{1,2} and practical \cite{pan} applications.
Among  the properties of the NLSE is the support of solitons as classical solutions. 
In the continuous case, the energy obtained from the hamiltonian of the NLSE   relates to the geometry of a curve by a Hasimoto transformation.
Thus, the corresponding solitons describe the buckling of the curve.
Similarly,   a properly discretized version of the NLSE preserves the integrability \cite{A,Iz} and supports soliton solutions. These 
can be utilized to model the buckling of piecewise linear polygonal chains, via a discretized version of the Hasimoto transformation \cite{IYN}.

In \cite{IYN}, a continuous string in $\mathbb R^3$ is represented by a two component complex spinor. The dreibein Frenet equation becomes a two component spinor Frenet equation which relates to the integrable hierarchy of the NLSE. That way the conserved charges of the NLSE hierarchy can be considered as hamiltonians; and can be  used  to  compute the energy  and govern the time evolution of a string. Finally, it has been  shown that the string hamiltonians in the NLSE hierarchy can be obtained alternatively by using a formalism of projection operators, originally introduced in the $\mathbb C \mathbb P^N$ models \cite{wojtek}.
In particular,   the connection between the projection operator formalism of the $\mathbb C \mathbb P^1$ model and the spinor Frenet equation has been explicitly  presented.

In \cite{IN,SYN,IYN} the Frenet frame formalism of \cite{IYN,Hu-2011} has been  generalized 
to the case of polygonal strings that are piecewise linear.
In particular, the Poisson geometry of a discrete string in $\mathbb R^3$  has been  derived. 
To do so,  the  Frenet frames are  converted  into a  spinorial representation,  the discrete
spinor Frenet equation is interpreted in terms of a  transfer matrix  formalism, and the Poisson brackets are determined in terms  of the spinor components.

The present article combines the results of \cite{IYN} and \cite{IN}  to study the  discrete nonlinear Schr\"odinger equation (DNLSE), its mathematical structure and its use in modelling discrete, piecewise linear filamental chains and strings.
In particular,  a bi-hamiltonian description of the DNLSE,  based on the discrete spinorial Frenet frame, is presented.

\section{The Discrete Frenet Frame}

Our starting point is the description of a discrete string in terms of an open and oriented, 
piecewise linear polygonal chain
${\Vr}(s) \in \mathbb R^3$ \cite{Hu-2011}. The {\it arc length}  parameter  takes values on 
$s \in [0,L]$ where $L$ is the total length of the string. 
The  vertices ${\mathscr D}_i$ (for $i \in  {\mathbb  Z}^+$) that specify the string  are located at the points
${\Vr}_i = ({\Vr}_0, \dots , {\Vr}_n)$  with ${\Vr}(s_i) = {\Vr}_i$; while,
the  endpoints are  ${\Vr}(0) = {\Vr}_0$ and  ${\Vr}(L) = {\Vr}_n$.
In addition, the distance of    the nearest neighbour vertices  ${\mathscr D}_{i}$ and ${\mathscr D}_{i+1}$  is
\[
|\Vr_{i+1} - \Vr_i | = s_{i+1} - s_i,
\]
which are connected by the {\it line segments}
\[
{\Vr}(s) \ = \ \frac{ s-s_{i} } {s_{i+1} - s_i} \, {\Vr}_{i+1}  \ - \ \frac{ s - s_{i+1}}  {s_{i+1} - s_i} \, {\Vr}_{i} ,
\ \ \ \ \ \ s\in(s_{i}, s_{i+1}),
\]
 and are characterised by  the unit length discrete tangent vector $\vt_i = (t_1, t_2, t_3)_i$ 
that points from  ${\mathscr D}_{i}$ to  ${\mathscr D}_{i+1}$ 
\be
\vec{t}_i=\fr{\vec{r}_{i+1}-\vec{r}_i}{|\vec{r}_{i+1}-\vec{r}_i|}.
\label{defti}\ee
Then, the discrete Frenet frame at the vertex ${\mathscr D}_{i}$ at the point $\vec{r}_i$ can be introduced assuming that the vertices at the points $\vec{r}_{i+1}$, $\vec{r}_i$ and $\vec{r}_{i+1}$ are not located on a common line (that is,  $\vt_i$ and $\vt_{i-1}$ are not parallel): 
The unit binormal vector is defined by
$$
\vecb_i=\fr{\vt_{i-1}\times \vt_i}{|\vt_{i-1}\times \vt_i|}
$$
while the unit normal vector $\vecn_i=\vecb_i\times \vt_i$ takes the form
$$
\vecn_i=\fr{-\vt_{i-1}+\left(\vt_{i-1}\cdot \vt_{i}\right)\vt_i\ }{|\vt_{i-1}+\left(\vt_{i-1} \cdot \vt_{i}\right)\vt_i|}.
$$
The orthogonal triplet $(\vec{t},\vec{n},\vec{b})_i$ constitutes the discrete Frenet frame associated the vertex ${\mathscr D}_{i}$ for each $i\in[1,n-1]$. 
The aforementioned formulae  are the two-point finite difference variants of the (corresponding) continuous Frenet frames. 
That way, the discretised curvature and torsion become geometrical.
They correspond to the {\it bond} and {\it torsion angle} which define the transfer angle that maps the discrete Frenet frame at  ${\mathscr D}_{i}$ to  ${\mathscr D}_{i+1}$ through the {\it transfer matrix} ${\cal R}_{i+1,i}$  
\begin{eqnarray}
\left(\begin{matrix} \vecn\\ \vecb \\ \vt \end{matrix} \right)_{i+1}
&=& {\cal R}_{i+1,i} \left(\begin{matrix} \vecn\\ \vecb \\ \vt \end{matrix} \right)_i
\nonumber\\
&=&\left(\begin{matrix}  \cos \bt \cos \bp & \cos \bt \sin \bp &-\sin \bt\\
-\sin \bp & \cos \bp &0\\
\sin \bt \cos\bp& \sin \bt \sin \bp &\cos \bt\\
\end{matrix} \right)_{i+1}
\left(\begin{matrix} \vecn\\ \vecb \\ \vt \end{matrix} \right)_i.
\label{DF}
\end{eqnarray}
Note that, the transfer matrix   is parametrized in terms of the two Euler angles  $(\bt,\bp)_i$  and represents an orthogonal rotation between the two consecutive discrete Frenet frames; i.e.,   is an element of $SO(3)$ as illustrated in Figure [1]. The third Euler angle has been excluded due to the orthogonality of $\vec{b}_i$ with $\vt_{i-1}$.

\begin{figure}[h]
        \centering
                \includegraphics[width=0.5\textwidth]{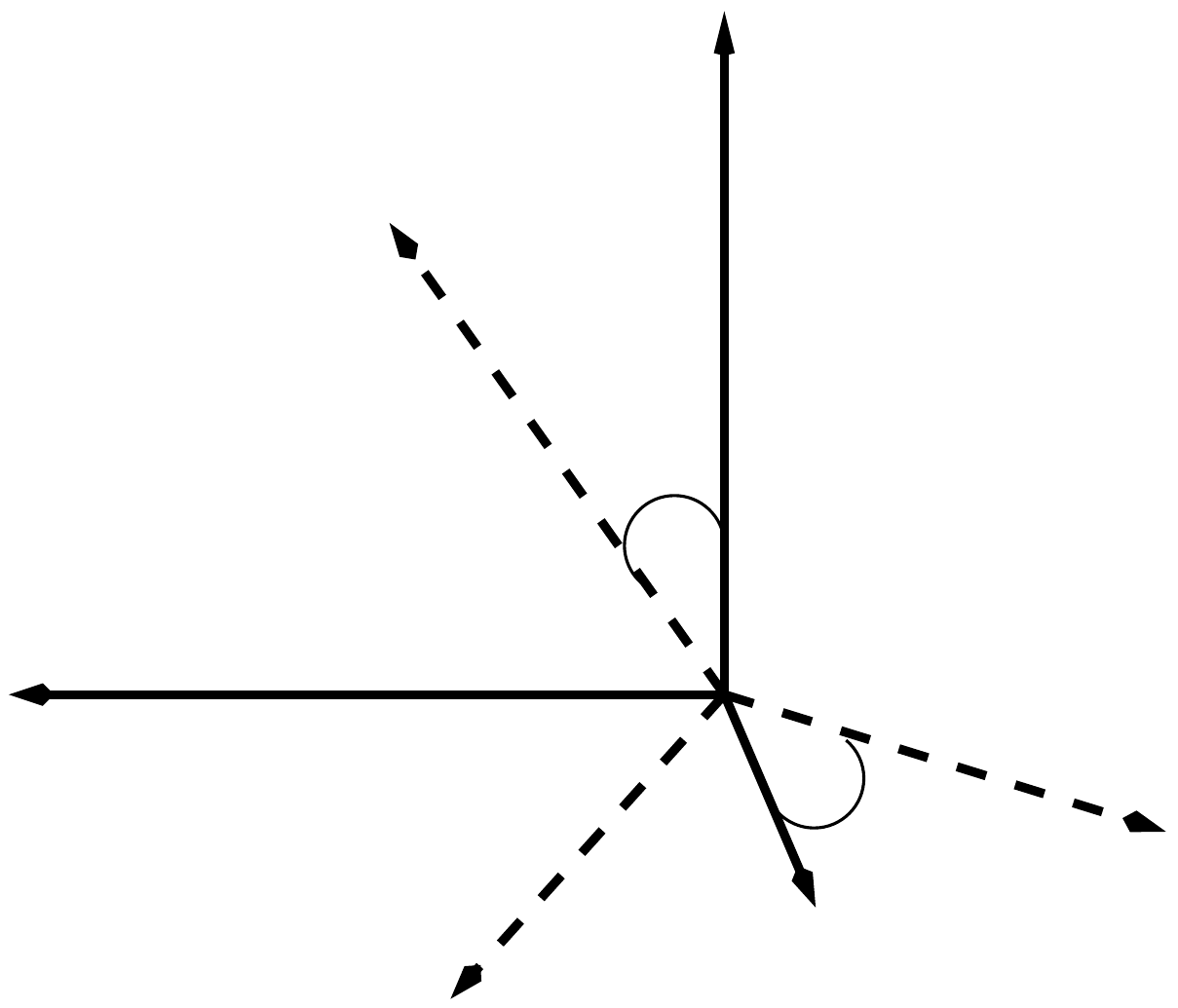}
           \put(10,30){$\vec{b}_{i+1}$}
             \put(-90,205){$\vec{t}_{i}$}
              \put(-250,75){$\vec{n}_{i}$}
                            \put(-95,25){$\vec{b}_{i}$}
              \put(-170,-10){$\vec{n}_{i+1}$}
              \put(-180,180){$\vec{t}_{i+1}$}
             \put(-70,35){$\bp_{i+1}$}
                          \put(-123,120){$\bt_{i+1}$}
        \caption{{ \it The bond and torsion angles of the transfer matrix ${\cal R}_{i+1}$.}}
       \label{Figure 2}
\end{figure}
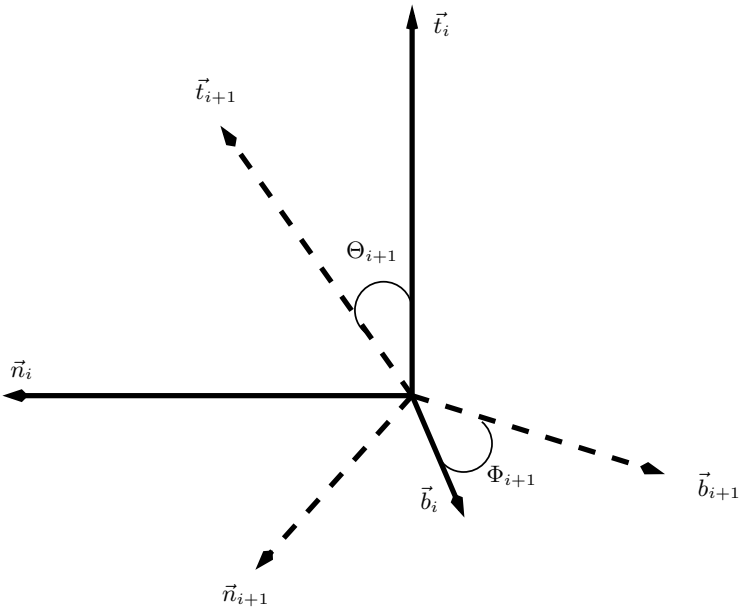

From  the transfer matrix  (\ref{DF}) all the discrete Frenet frames can be constructed since 
\be
\vt_{i+1}=\sin \bt_{i+1} \cos \bp_{i+1} \,\vecn_i +\sin \bt_{i+1} \sin \bp_{i+1} \,\vecb_i+
\cos \bt_{i+1}\,\vt_i\nonumber
\ee
implying that   
\be 
\vt_{i+1}\cdot \vt_{i}&=& \cos \bt_{i+1} \label{eq1}\nonumber\\
\vt_{i+1}\cdot \vecn_i&=&\sin \bt_{i+1}\, \cos \bp_{i+1}\nonumber\\
\vt_{i+1}\cdot \vecb_i&=&\sin \bt_{i+1}\, \sin \bp_{i+1}.
\label{eq1}
\ee
Similarly, from the other rows  one gets
\be
\vecb_{i+1}\cdot \vecb_i&=&\cos \bp_{i+1}\nonumber\\
\vecb_{i+1}\cdot \vecn_i&=&-\sin \bp_{i+1}\nonumber\\
\vecb_{i+1}\cdot \vt_i&=&0,
\label{eq2}
\ee
and 
\be 
\vecn_{i+1}\cdot \vt_{i}&=& -\sin \bt_{i+1} \nonumber\\
\vecn_{i+1}\cdot \vecb_i&=&\cos \bt_{i+1}\, \sin \bp_{i+1}\nonumber\\
\vecn_{i+1}\cdot \vecn_i&=&\cos \bt_{i+1}\, \cos \bp_{i+1}.
\label{eq3}
\ee
Thus,   $\Theta_{i+1}$ is the bond angle and $\Phi_{i+1}$ is the torsion angle corresponding to the discrete analogue of the continuum curvature and torsion, respectively; i.e.,  the chain is determined by its bond and torsion angles uniquely up to global rotation and translation.

\section{ The discrete spinor Frenet equation}

In \cite{IN},  an alternative approach to represent discrete piecewise linear 
chains is introduced and developed to describe  their time evolution, and the two dimensional Riemann surfaces that are swept by  time evolution.
In particular, the spinorial formulation of a discrete string in combination with the formalism of the discrete  Frenet equations has been employed as described below.

Each link from vertex ${\mathscr D}_i$ to vertex  ${\mathscr D}_{i+1}$ is associated to
 a two component {\it complex spinor }
\begin{equation}
\psi_{i} =  \left( \begin{matrix} z_{1}  \\ z_{2} \end{matrix}\right)_i,
\label{psi1}
\end{equation}
where $z^{i}_{a}$ (for $a=1,2$) are complex variables with support 
on the {\it link}. 
The spinors are related to the unit length tangent vectors by the relation
\[
\sqrt{g_i}\, \vec{ t}_i  = <\psi_i, \hat \sigma \psi_i> 
\]
where $\vec{\sigma} = (\sigma^1, \sigma^2, \sigma^3)$ are the Pauli matrices and
\[
\sqrt{g_i} \equiv  |z^{i}_{1}|^2 + |z^{i}_{2 }|^2
\]
is a metric scale factor. 
Explicitly,
\begin{equation}
\left( \begin{matrix} t^i_1+ i t^i_2  \\ t^i_3 \end{matrix}\right) \ = \ \fr{1}{\sqrt{g_i} }
\left( \begin{matrix} 2 \bar z^{i}_{1}  z^{i}_{2} \\  
| z^{i}_{1}|^2 
- |z^{i}_{2}|^2  \end{matrix} \right).
\label{tz1z2}
\end{equation}
Together with (\ref{defti}) this determines the spinor components $z_\alpha^i$ in terms of the vertices $\mathcal D_i$, up to an overall phase.
In addition, for each $i$ the conjugate spinor $\bar\psi_i$ is  defined by introducing 
the {\it charge conjugation} operation ${\mathscr C}\,$ that acts on $\psi_i$ in the following way
\begin{equation}
{\mathscr C} \,  \psi_i \ = \  -i \sigma_2 \psi_i^\star = \bar \psi_i \ = \ \left( \begin{matrix} - \bar z_{2} 
\\ \  \ \bar z_{1} \end{matrix}\right)_i.
\label{psi2}
\end{equation}
Since  $ \mathscr C^2 = - \mathbb I$ the spinors are orthogonal by construction, i.e. $<\psi_i \, , \bar\psi_i> = 0$. In addition, they are combined through   the  $SU(2)$ matrix 
\[
u_i = \left( \begin{matrix} z_1 & -\bar z_2 \\ z_2 & \ \  \bar  z_1 \end{matrix} \right)_i
\]
since 
\[
\psi _i  =   {\mathfrak u}_i  
\left( \begin{matrix} 1 \\ 0 \end{matrix} \right) \ \ \ \ \& \ \ \ \  \bar \psi_i  =   {\mathfrak u}_i  
\left( \begin{matrix} 0 \\ 1 \end{matrix} \right)
\]
For completeness, note that 
\[
u_i\sigma_3u_i^{-1}=(\vec{t}\cdot \vec{\sigma})_i
\]  
which is in agreement with equation (\ref{tz1z2}).

In the continuum case, the parametrisation of the {\it link variables} $z_\alpha^i$ in terms of the local angular variables $\ct$ and $\cp_{\pm}=\cp_2\pm\cp_1$  gives rise to the classical version of ${\it sl}_2$ which appears in the case of local spin chains  \cite{IYN}. 
Its discrete analogue  is of the form
\be
z_1^{i}=\cos \fr{\ct}{2} \, e^{-\fr{i\cp}{2}}\large{|}_{i} \ \ \ \ \ \&  \ \ \ \ \ z_2^{i}=\sin \fr{\ct}{2} \, e^{\fr{i\cp}{2}}\large{|}_{i}
\label{z1z2}
\ee
assuming that  $\cp_+=0$ and $\cp_{-}\equiv\cp$ (due to the overall phase).
Note that, in the  local coordinate representation (\ref{z1z2}) the metric equals one.

In  \cite{IN}, it was shown that  by combining the two spinors  (\ref{psi1}) and (\ref{psi2}) into a  Majorana spinor
\begin{equation*}
\Psi_i = \left( \begin{matrix}  -\bar\psi\\  \ \ \psi\end{matrix} \right)_i,
\label{majspi}
\end{equation*}
 the discrete Frenet euqation (\ref{DF}) is equivalent to the spinorial one
\begin{equation}
\Psi_{i+1} \ = \  \mathcal U_{i+1}^\dagger \ \Psi_i.\label{discfre1}
\end{equation}
 $\mathcal U_{i} $ is the {\it ensuing transfer matrix} in the chain of spinors $\Psi_i$ defined by the {\it rotational variables} $Z^{i}_{a}$ (for $a=1,2$)
\begin{equation*}
\mathcal U_{i} \ = \ \left( \begin{matrix}  Z_{1} 
& - \bar Z_{2}\\   Z_{2}& \ \ \bar Z_{1}\end{matrix} \right)_i
\label{matU}
\end{equation*}
where
\be Z_1^{i}=\cos \fr{\bt_i}{2} \, e^{-\fr{i\bp_i}{2}} \ \ \ \ \ \ \& \ \ \ \ \ Z_2^{i}=\sin \fr{\bt_i}{2} \, e^{\fr{i\bp_i}{2}}.
\label{bZ}
\ee 
It is straightforward to see that  
\[
\cos\bt_{i}=\left(2|Z_2^{i}|^2-1\right) \ \ \ \ \ \ \& \ \ \ \ \  \cos\bp_{i}=\fr{1}{2}\left\{\left[(Z_1^{i})^2+(\bar Z_2^{i})^2\right]+\mbox{cc}\right\}
\] 
in accordance with the Frenet frame obtained from  equations (\ref{eq1}) and (\ref{eq2}). For example, 
\[
\vt_{i+1}\cdot\vt_{i}=\left(2|Z_2^{i}|^2-1\right)
\] 
and so on.

Finally, the relation  between the link $z_a^i$ and the rotational variables  $Z_a^i$  is given  by (\ref{discfre1}) since
\begin{equation*}
\mathcal U_{i+1}^\dagger=\fr{1}{\sqrt{g_i}}\left(\Psi_{i+1}\Psi_i^\dagger\right)
\label{deftU}
\end{equation*}
 leading to 
\begin{eqnarray}
 Z_{1}^{i} &=&\fr{1}{\sqrt{g_i}} \left(\bar z_{1}^{i+1} z_{1}^{i} + \bar z_{2}^{i+1} z_{2}^{i}\right)  \nn \\ 
 Z_{2}^{i} &=& \fr{1}{\sqrt{g_i}} \left(z_{1}^{i+1} z_{2}^{i}  - z_{2}^{i+1} z_{1}^{i}\right).
\label{Z}
\end{eqnarray}

Let us conclude, by discussing the discrete analogue of the equivalence between the projection operator formalism of the $\mathbb C \mathbb P^1$ model  and  the classic Frenet equation of a continuous string \cite{IYN}. To do so, let us define the  discrete projection operator 
\begin{equation}
\mathbb P_i \ = \   \psi_i \!\otimes \! \psi_i^\dagger \ = \ u_i \left( \begin{matrix} 1 & 0 \\ 0 & 0 
\end{matrix} \right) u_i^{-1}
\label{P}
\end{equation}
accompanied by the complemental projection operator 
\begin{equation}
\bar{\mathbb P}_i \ = \ \bar{\psi}_i \!\otimes \! \bar{\psi}_i^\dagger \ = \ u_i \left( \begin{matrix} 0 & 0 \\ 0 & 1 
\end{matrix} \right) u_i^{-1}
\label{P}
\end{equation}
such that  the following relations exist
\[
\mathbb P_i+\bar{\mathbb P}_i=\mathbb I_i  \ \ \ \ \ \ \& \ \ \ \ \  \mathbb P_i^2=\mathbb P_i
\]
and so forth;  for details, see \cite{IYN}.  Note that, the identity 
\[
\mathbb P_i^2=\mathbb P_i
\] 
sets the metric equal to one; i.e.,  $\sqrt{g_i}=1$  in accordance with  the local  angular parametrization (\ref{z1z2}). 
This was not the case in \cite{IN} where the Poisson algebra of the link $z_\alpha^i$and rotational $Z_\alpha^i$ variables have been presented due to a different (than   the projection operator formalism) approach.
 Therefore,  the Poisson brackets of the corresponding  link variables $z_\alpha^i$ (given in  Appendix B) are different from the ones presented in \cite{IN}.
 
Finally, using  the fact that the elements of the Frenet frame are $su(2)$ Lie algebra generators  the following equivalence has been derived 
 \be
 \mathbb P_i-\bar{\mathbb P}_i\simeq \vec{t}_i.\label{Pt}\ee
 In the continuum \cite{IYN}, it was shown    that the time evolution of the spinor (that describes the string) can be obtained from  the integrable hierarchy of NLSE. 
  In addition, it was shown that the conserved charges of the NLSE hierarchy can be expressed in terms of the projection operator formalism and, in particular,  that are equivalent to the  energy density of the  $\mathbb C \mathbb P^1$ chiral model.

In the next section, we use the DSLNE to introduce an energy function for discrete polygonal chains. 
Equation (\ref{discfre1}) implies that the rotational angles are the natural variables for constructing energy functions for discrete piecewise linear strings.
In analogy, with the continuum case, the energy must remain invariant under the local $SO(2)$ frame rotation.
For this we simply identify the spin variables with the tangent vector  in accordance with  (\ref{P}). 
Thus, the energy function determines a chain dynamics which is manifestly invariant under the local frame rotation.

\section{Bi-Hamiltonian Structure}

In order to study the dynamics of a curve, Hamilton's equation of motion is a natural choice.
In \cite{IYN}, using the projection formalism of the $\mathbb C \mathbb P^1$ model,  it was shown that  the string hamiltonians in the NLSE hierarchy are related to the spinor Frenet equation. 

In fact, it was shown that  the continuous string hamiltonian  via the projection operator  formalism 
 \be{\cal H}^1_{\tiny{\mbox{NLSE}}}= \mbox{tr}\left(\mathbb P_s\,\mathbb P_s\right)
\label{HP}\ee 
due to (\ref{Pt}), is equivalent to  the number operator in the NLSE hierarchy (otherwise, the Heisenberg spin chain hamiltonian)
\be{\cal H}^2_{\tiny{\mbox{NLSE}}}=\fr{1}{2} \left(\pr_s \vt \cdot \pr_s \vt\right).\label{Ht}
\ee
That is,
\be
{\cal H}_{\tiny{\mbox{NLSE}}}={\cal H}^1_{\tiny{\mbox{NLSE}}}={\cal H}^2_{\tiny{\mbox{NLSE}}}=\fr{1}{2}|\kappa_c|^2,
\label{cH}
\ee
where $\kappa_c$ is the complex valued curvature (for details, see \cite{IYN}).

In the discrete case, the polygonal curve is a set of sequentially connected segments whose length here is considered to be locally inextensible and therefore,  the dynamics of the  curve is  essential the rotation of tangent vectors.
Since the degree of freedom for tangent vectors is the angular rotation, the chain can be identified as  a spin chain  described by the lattice isotropic Heisenberg magnet model. 

Thus, for discretising ${\cal H}_{\tiny{\mbox{NLSE}}}$ given by (\ref{cH}) the forward finite difference scheme  (i.e.,  $\pr_s f=f_{i+1}-f_i$) is  applied. That way, one  obtains
\be
\mathcal H_{\tiny{\mbox{DNLS}}}={\cal H}^1_{\tiny{\mbox{DNLSE}}}={\cal H}^2_{\tiny{\mbox{DNLSE}}}.
\label{gH}\ee

The first hamiltonian ${\cal H}^1_{\tiny{\mbox{DNLSE}}}$ is the discretized version of the $\mathbb C \mathbb P^1$ sigma model (\ref{HP})  given by 
\begin{eqnarray}
{\cal H}^1_{\tiny{\mbox{DNLSE}}}&=&\,\sum_i \mbox{tr}\left(\mathbb P_{i+1}-\mathbb P_i\right)^2\nonumber\\
&=&2\sum_i |z_1^iz_2^{i+1}-z_1^{i+1}z_2^i|^2\label{Hz}
\end{eqnarray}
provided that  $\sqrt{g_i}=1$. 

 The second hamiltonian ${\cal H}^2_{\tiny{\mbox{DNLSE}}}$ is the discretized version of NLSE hierarchy (\ref{Ht})  given by 
\begin{eqnarray}
{ \cal H}^2_{\tiny{\mbox{DNLSE}}}&=&\sum_i \left(1-\vt_{i+1}\cdot \vt_{i}\right)\nonumber\\
&=&2 \sum_i  \,|Z_2^i|^2\label{HZ}
\end{eqnarray}
that is, the  isotropic (XXX) Heisenberg model hamiltonian which describes  first neighbour   spin-spin interactions. 
Note that,  equation (\ref{gH}) holds   due to (\ref{Z}). 

Finally, the equation of motion is given by
\begin{eqnarray*}
\fr{d\vt_i}{dt}&=&\left\{\mathcal H_{\tiny{\mbox{DNLS}}},\vt_i\right\}\\
&=&\vt_i\times \fr{\pr \mathcal H_{\tiny{\mbox{DNLS}}}}{\pr \vt_i}.
\end{eqnarray*}
In what follows, two a priori different discrete versions of the continuum NLSE are presented. 
Recall that, the NLSE equation is known to be integrable at the finite segment $[0,L]$  when appropriate boundary conditions are imposed. 
The aforementioned integrability is extended in the discretized version since a bi-hamiltonian form of (\ref{Hz}) is presented. 

The first hamiltonian  (\ref{Hz}) is given in terms of the local angular variables (\ref{z1z2}), that is,
 \be 
 \mathcal H^1_{\tiny{\mbox{DNLS}}}= \sum_i\left[1-\cos \ct_{i+1}\cos \ct_{i}-\sin \ct_{i+1}\sin \ct_{i}\cos (\cp_{i+1}-\cp_i)\right]
 \label{H1}
 \ee
 associated with the only {\it non-zero} Poisson brackets:
 \be 
\left\{ \cos \ct_i,\cp_j \right\}=\delta_{ij}.
\label{cc}
 \ee 
In terms of the tangent vectors $\vt_i$, the Poisson brackets are given by $\{t_i^a,t_j^b\}=-\varepsilon^{abc} t_i^c\delta_{ij}$. The consistency of the  Poisson brackets between the two systems of variables is discussed  in Appendix A.
 
 Next, the second  hamiltonian (\ref{HZ}) in terms of the discrete Euler's angles  (\ref{bZ}) is presented.
However,  in order to obtain  the corresponding Poisson brackets  the relation between the two angular systems need to be known explicitly.
In particular, from (\ref{Z}) and  (\ref{z1z2}) it is straightforward to   obtain that 
 \be
 \cos\bt_{i+1}&=&\left(2|Z_1^{i+1}|^2-1\right)\nonumber\\
 &=&\cos\ct_{i+1}\cos\ct_{i}+\sin\ct_{i+1}\sin\ct_{i}\cos \left(\cp_{i+1}-\cp_i \right)
 \label{ct}\acc
  \sin\bt_{i+1}&=&2Z_1^{i+1}Z_2^{i+1}\nonumber\\
   &=&\sin\ct_{i+1}\cos\ct_i-\cos\ct_{i+1}\sin\ct_i \cos\left(\cp_{i+1}-\cp_i\right)\label{st} \acc
 \cos\bp_{i+1}&=&\fr{1}{2}\left\{\left[(Z_1^{i+1})^2+(\bar Z_2^{i+1})^2\right]+\mbox{cc}\right\}\nonumber \\
 &=&\cos\left(\cp_{i+1}-\cp_i\right)\label{cp}\acc
  \sin\bp_{i+1}&=&\fr{i}{2}\left\{\left[(Z_1^{i+1})^2+(\bar Z_2^{i+1})^2\right]-\mbox{cc} \right\}\nonumber\\
  &=&\cos\ct_{i}\sin\left(\cp_{i+1}-\cp_i\right),\label{sp}\ee
  associated with the mixed terms
  \be 
 \cos\bt_{i+1} \cos\bp_{i+1}&=&\fr{1}{2}\left\{\left[(Z_1^{i+1})^2-(\bar Z_2^{i+1})^2\right]+\mbox{cc}\right\} \nonumber\\
   &=&\cos\ct_{i+1}\cos\ct_i \cos\left(\cp_{i+1}-\cp_i\right)+\sin\ct_{i+1}\sin\ct_i \label{ctcp}\acc
 \cos\bt_{i+1} \sin\bp_{i+1}&=&\fr{i}{2}\left\{\left[(Z_1^{i+1})^2-(\bar Z_2^{i+1})^2\right]-\mbox{cc} \right\}\nonumber\\
 &=&\cos\ct_{i+1} \sin\left(\cp_{i+1}-\cp_i\right)\label{ctsp}\acc
 \sin\bt_{i+1} \cos\bp_{i+1}&=&Z_1^{i+1} \bar Z_2^{i+1}+\mbox{cc}\nonumber\\
 &=& -\cos\ct_{i+1}\sin\ct_i+\sin\ct_{i+1}\cos\ct_i \cos\left(\cp_{i+1}-\cp_i\right)\label{stcp} \acc
 \sin\bt_{i+1} \sin\bp_{i+1}&=&i\left(Z_1^{i+1} \bar Z_2^{i+1}-\mbox{cc} \right)\nonumber\\
 &=&\sin\ct_{i+1}\sin\left(\cp_{i+1}-\cp_i\right).
 \label{stsp}
 \ee
 Also, due to consistency, the following condition is  satisfied
 \be
 \sin\ct_i\sin\left(\cp_{i+1}-\cp_i\right)=0
 \label{prop}
 \ee
 which follows from the   trigonometric identity $\sin^2\bp_{i+1}+\cos^2\bp_{i+1}=1$ using equations  (\ref{ct}) and (\ref{cp}).
 Also, from  (\ref{Z}), (\ref{prop}) and (\ref{ten}), it is straightforward to show the equivalence between the parametrization (\ref{eq1})-(\ref{eq3}) and the equations (\ref{ct})-(\ref{stsp}).

So, in terms of the discrete Euler's angles, the second hamiltonian (\ref{HZ}) takes the simple form
 \be 
\mathcal H^2_{\tiny{\mbox{DNLS}}}= \sum_i\left(1-\cos \bt_{i+1}\right).
 \label{Hd2}
 \ee
associated with the Poisson brackets
\be
\{\bt_{i+1},\bt_i\}&=&-\sin\bp_{i+1}\label{1}\\\
\{\bp_{i+1},\bp_i\}&=&\fr{1}{\sin\bt_i}\left(\sin\bp_{i+1}\cot\bt_{i+1}+\sin\bp_i\cot\bt_{i-1}\right)\label{2}\\
\{\bt_{i+1},\bp_i\}&=&\fr{\cos\bp_{i+1}}{\sin\bt_i}\label{3}\\
\{\bt_{i},\bp_i\}&=&-\cot\fr{\bt_i}{2}-\cos\bp_i\cot\bt_{i-1}\label{4}\\
\{\bt_{i-1},\bp_i\}&=&\cot\fr{\bt_{i-1}}{2}+\cos\bp_i\cot\bt_{i}\label{5}\\
\{\bt_{i-2},\bp_i\}&=&-\fr{\cos\bp_{i-1}}{\sin\bt_{i-1}}
\label{6}
\ee
in addition with the shifted ones to the left and right; i.e.,  $i \rightarrow i-1$ and  $i \rightarrow i+1$ respectively.  For completeness, in  Appendix C the consistence of the two algebras (\ref{cc}) and (\ref{1})-(\ref{6}) is presented. 
 Note that, the Poisson brackets (\ref{1})-(\ref{6}) was first introduce  in \cite{SYN}  but in another context.

Let us conclude by stating that two different hamiltonians for the DNLS have been obtained. The first $\mathcal H^1_{\tiny{\mbox{DNLS}}}$   (\ref{H1}) has a  "complicate" form
associated by a "simple algebra" (\ref{cc}) while the second hamiltonian $\mathcal H^2_{\tiny{\mbox{DNLS}}}$ (\ref{Hd2}) has a "simple" form associated with a "complicate"  algebra (\ref{1})-(\ref{6}).

 \section{Conclusion}

In summary, we have addressed the problem how to utilize extrinsic geometry in order to derive hamiltonian energy functions for discrete strings that move in three dimensional space.
We utilize the transfer matrix formalism to consecutively map the discrete Frenet frame from one vertex of discrete curve to its neighbour.
In this manner we arrive naturally to energy densities that relate to the integrable hierarchy of the NLSE defined on the lattice. 
In particular, under this frame the tangent vectors are considered to be classical spins and thus, the lattice Heisenberg model comes to be a natural choice of the chain's hamiltonian.

We have shown that the introduced discrete version of the NLSE is integrable since it holds a bi-hamiltonian structure. 
That way, the equation of motion of the tangent vectors and bond and torsion angles Poisson brackets are constructed and therefore, the dynamics of the lattice chain can be investigated. 
We will report on these issues in a future publication.

\section{Acknowledgments}
T.I. thanks  the School  of Physics at Beijing Institute of Technology  for hospitality during the completion of this research. 

T.I. acknowledges  support from FP7, Marie Curie Actions, People, International Research Staff Exchange Scheme (IRSES-606096); and  from The Hellenic Ministry of Education: Education and Lifelong Learning Affairs, and European Social Fund: NSRF 2007-2013, Aristeia (Excellence) II (TS-3647).
A.J.N.  acknowledges  support from CNRS PEPS grant, Region Centre 
Rech\-erche d$^{\prime}$Initiative Academique grant; Sino-French Cai Yuanpei Exchange Program (Partenariat Hubert Curien), Vetenskapsr\aa det, Carl Trygger's Stiftelse f\"or vetenskaplig forskning; and  Qian Ren Grant at BIT.

\appendix
\section{}
In the discrete case, the unit length tangent vector $\vt=\fr{1}{\sqrt{g_i}}\left( {\psi}^\dagger\vec{\sigma}\psi\right)$  and the  binormal and  normal vectors $\vn+i\vec{b}=\fr{1}{\sqrt{g_i}}\left(\bar{\psi}^\dagger\vec{\sigma}\psi\right)$  in terms of the bond and torsion angles (\ref{z1z2}) are given by
\be
\vt_{i}=\fr{1}{\sqrt{g_i}}\left(\br \bar{z}_1z_2+\bar{z_2}z_1\\ i(\bar{z}_2z_1-\bar{z}_1z_2)\\|z_1|^2-|z_2|^2\er\right)_{i}=\left(\br \sin \ct \cos \cp\\ \sin \ct \sin \cp\\ \cos \ct \er\right)_{i}
\label{t}\ \ \ \ \ \ 
\vecn_{i+1}=\left(\br \cos \ct \cos \cp \\ \cos \ct \sin \cp \\ -\sin \ct\er\right)_{i+1},\ \ \ \ \ 
\vecb_{i+1}=\left(\br -\sin  \cp \\ \ \ \cos \cp \\ 0\er\right)_{i+1}\label{ten},
\ee
where $\psi$ is given by (\ref{psi1}) and $\sqrt{g_i}=1$.
Then  the continuum hamiltonian (\ref{cH}), using the forward finite difference scheme,
becomes
\be \mathcal H_{\mbox{\tiny{NLSE}}}=2\sum_{i}(1-\vec{t}_{i+1}\cdot \vec{t}_i)
\ee 
accompanied by the Poisson bracket 
\be
\left\{t_i^a,t_j^b\right\}=-\varepsilon^{abc} \,t_i^c\,\delta_{ij}
\label{po}
\ee
in consistence with (\ref{Hz}). 

Another way to obtain (\ref{po}) is to use the continuum XXX chain formulation.
That is, assume that the  canonical coordinates are $p_i=\cos \ct_i$ and $q_i=\cp_i$ where $\{p_i,q_j\}=\delta_{ij}$ and $\{q_i,q_j\}=0=\{p_i,q_j\}$ (in consistence with  (\ref{cc})). Also, assume  that  $t_i^\pm=t_i^1\pm i t_i^2=\sin \ct_i \,e^{\pm i\phi_i}$ and $t_i^z=t_i^3=\cos \ct_i$ due to  (\ref{t}). 
Thus, in  terms of the canonical coordinates this implies that $t_i^\pm=\sqrt{1-p_i^2}\,e^{\pm iq_i}$ and $t_i^z=p_i$ and so it is easy to obtain that 
\be
\{t_i^+,t_j^-\}=-2it_i^z\delta_{ij},\  \  \ \ \ \ \ \{t_i^z,t_j^\pm\}=\mp i \,t^\pm \, \delta_{ij}
\ee
that is,   (\ref{po}).

\section{}
In terms of the link variables $z_a^i$ variables  the associated  Poisson brackets are non-trivial. In fact, from  (\ref{z1z2})  using (\ref{cc}) it can be shown that the only non-vanishing Poisson brackets  are the following:
 \begin{eqnarray*}
 \{z_a^i,\bar{z}^j_a\}&=&\fr{i}{4}\,\delta_{ij}, \\
  \{z_1^i,z_2^j\}&=&-\fr{i}{8}\left(\fr{z_1^i}{\bar{z}_2^i}-\fr{z_2^i}{\bar{z}_1^i}\right)\delta_{ij}\\
  \{z_1^i,\bar{z}_2^j\}&=&-\fr{i}{8}\left(\fr{z_1^i}{z_2^i}+\fr{\bar{z}_2^i}{\bar{z}_1^i}\right)\delta_{ij}.
  \end{eqnarray*}
Recall  that, $|z_1^i|^2+|z_2^i|^2=1$.

\section{}
Let us conclude by presented some explicit examples of how the Poisson brackets of the Euler's angles $(\bt,\bp)_i$  can be obtained directly from the corresponding ones of the local angular variables $(\ct,\cp)_i$. 

Let us start by considering the first bracket (\ref{1}). 
Substituting  equation (\ref{ct}) in the right hand side of (\ref{1})  and applying the properties of the Poisson bracket one gets
\be 
\{\cos\bt_{i+1},\cos\bt_i\}&=&\fr{\pr}{\pr\ct_i}\left(\cos\bt_{i+1}\right)\,\{\ct_i,\cp_i\}\,\fr{\pr}{\pr\cp_i}\left(\cos\bt_{i}\right)-\fr{\pr}{\pr\cp_i}\left(\cos\bt_{i+1}\right)\,\{\ct_i,\cp_i\}\,\fr{\pr}{\pr\ct_i}\left(\cos\bt_{i}\right)\nonumber\\
&=&\sin\bt_{i+1}\cos\bp_{i+1}\left(\fr{-1}{\sin\ct_i} \right)\left[-\sin\ct_i \sin\ct_{i-1}\cos(\cp_i-\cp_{i-1})\right]\nonumber\\
&-&
\sin\ct_{i+1}\sin\ct_i\sin(\cp_{i+1}-\cp_{i})\left(\fr{-1}{\sin\ct_i}\right)\left(-\sin\bt_i\right)\nonumber\\
&=&0-\sin\ct_{i+1}\sin(\cp_{i+1}-\cp_{i})\sin\bt_i\nonumber\acc
&=&-\sin\bt_{i+1}\sin\bp_{i+1}\sin\bt_i.\ee
Note that, the condition (\ref{prop}) and the equation (\ref{stsp}) is (also) been used. 

However, the right-hand side of the above equation gives
\be
\left\{\cos\bt_{i+1},\cos\bt_i\right\}&=&\sin\bt_{i+1}\sin\bt_i\,\{\bt_{i+1},\bt_i\}.\ee
Thus,  equation (\ref{1}) is obtained.

For completeness, let us (also) show how the Poisson bracket (\ref{3}) can be derived. 
Substituting  equation (\ref{ct}) and (\ref{sp}) in the right hand side of (\ref{3})  and applying the properties of the Poisson bracket one gets
\be 
\{\cos\bt_{i+1},\sin\bp_i\}&=&\fr{\pr}{\pr\ct_i}\left(\cos\bt_{i+1}\right)\,\{\ct_i,\cp_i\}\,\fr{\pr}{\pr\cp_i}\left(\sin\bp_{i}\right)-\fr{\pr}{\pr\cp_i}\left(\cos\bt_{i+1}\right)\,\{\ct_i,\cp_i\}\,\fr{\pr}{\pr\ct_i}\left(\sin\bp_{i}\right)\nonumber\\
&=&\sin\bt_{i+1}\cos\bp_{i+1}\left(\fr{-1}{\sin\ct_i}\right)\cos\ct_{i-1}\cos\left(\cp_i-\cp_{i-1}\right)\nonumber\\
&=&\sin\bt_{i+1}\cos\bp_{i+1}\left(\fr{-1}{\sin\ct_i}\right)\cos\ct_{i-1}\cos\bp_i\nonumber\acc
&=& \sin\bt_{i+1}\cos\bp_{i+1}\sin \bt_i \cos\bp_i.
\ee
Equation  (\ref{cp}) has (also) been used as well as the identity $\sin\bt_i\equiv\fr{\sin\ct_i}{\cos\ct_{i-1}}$ obtained  when  (\ref{stsp}) is divided by (\ref{sp}).

However, the right-hand side of the above equation gives
\be
\left\{\cos\bt_{i+1},\sin\bp_i\right\}&=&-\sin\bt_{i+1}\cos\bp_i\,\{\bt_{i+1},\bp_i\}.\ee
Therefore,  equation (\ref{3}) is derived.


\begin{thebibliography}{99}

\bibitem{vortex} P.G. Saffman, {\it Vortex Dynamics}  (Cambridge University Press, Cambridge, 1992)


\bibitem{pauls} N. Manton and P. Sutcliffe,  {\it Topological Solitons} (Cambridge University Press, Cambridge, 2004)

\bibitem{danielsson} U. Danielsson, M. Lundgren, A.J. Niemi, Phys. Rev. {\bf E82}  021910 (2010)

\bibitem{nora1} M. Chernodub, S. Hu, A.J. Niemi, Phys. Rev. {\bf E82} 011916 (2010) 

\bibitem{nora2} N. Molkenthin, S. Hu, A.J. Niemi  Phys. Rev. Lett. {\bf 106}  078102 (2011) 

\bibitem{hansonbook} A.J. Hanson, {\it Visualizing Quaternions} (Morgan Kaufmann Elsevier, London, 2006)

\bibitem{Has}
H. Hasimoto, J. Phys. Soc. Japan {\bf 31} 293 (1071); J. Fluid Mech. {\bf 51} 477 (1972)


\bibitem{1}
L.D. Faddeev and L.A. Takhtajan, {\it Hamiltonian Methods in the Theory of Solitons} (Springer Verlag, Berlin, 1987)

\bibitem{2}
 M.J. Ablowitz, B. Prinari and A.D. Trubatch, {\it Discrete and Continuous Nonlinear Schr\"odinger
Systems} (Cambridge University Press, Cambridge, 2004)

\bibitem{pan}
P.G. Kevrekidis, {\it The discrete Nonlinear Sch\"odinger equation: Mathematical Analysis, Numerical Computations and Physical Perspectives}  (Springer Verlag, Berlin, 2009)

\bibitem{A}
M.J. Ablowitz and J.F. Ladik, Stud. Appl. Math. {\bf 17} 1011 (1976)

\bibitem{Iz}
A.G. Izergin and V.E. Korepin,  Dokl. Akad. Nauk. {\bf 259} 76 (1981)

\bibitem{IYN} T. Ioannidou, Y. Jiang and  A.J. Niemi,  Phys. Rev. D {\bf 90} 025012  (2014)


\bibitem{SYN}
S. Hu, Y. Jiang and A.J. Niemi, Phys. Rev. D {\bf 87}  10 (2013) 

\bibitem {IN}
T. Ioannidou and  A.J. Niemi,  Phys. Lett. A {\bf 380} 333 (2016) 


\bibitem{Hu-2011} S. Hu, M. Lundgren and A.J. Niemi, Phys. Rev.  E {\bf 83}  061908 (2011)


\bibitem{wojtek} 
W.J. Zakrzewski, {\it Low Dimensional Sigma Models} (Institute of Physics Publishing, London, 1989)


\end{thebibliography}
\end{document}